# Hebbian control of fixations in a dyslexic reader

Albert Le Floch [a,b] , Guy Ropars [a,c*]

[a]*Laser Physics Laboratory, University of Rennes, 35042 Rennes Cedex, France*
[b]*Quantum Electronics and Chiralities Laboratory, 20 Square Marcel Bouget, 35700 Rennes, France*
[c]*UFR SPM, University of Rennes, 35042 Rennes Cedex, France*

*\* Corresponding author at: Laser Physics Laboratory, University of Rennes, 35042 Rennes Cedex, France. e-mail: guy.ropars@univ-rennes.fr ; albert.lefloch@laposte.net*

**During reading, dyslexic readers exhibit more and longer fixations than normal readers. However, there is no significant difference when dyslexic and control readers perform only visual tasks on a string of letters, showing the importance of cognitive processes in reading. This linguistic and cognitive processing demand in reading is often perturbed for dyslexic readers by perceived additional letter and word mirror-images superposed to the primary images on the primary cortex, inducing an internal visual crowding. Here we show that whereas for a normal reader, the number and the duration of fixations remain invariant whatever the nature of the lighting, the excess of fixations and total duration of reading can be controlled for a dyslexic reader using the Hebbian mechanisms to erase the extra images in an optimized pulse-width lighting. The number of fixations can be reduced by a factor of about 1.8, recovering the normal reader records.**

## Introduction

Fixational eye movements are a fundamental aspect of vision[1–3]. Even when a compound eye like that of drosophila is rigidly attached to the skeleton, muscles have recently been shown to move the retina itself [4]. For decades, anomalous eye movements have been discussed in particular in children and adults with dyslexia[2,5].

Binocular coordination of saccades[2,6], and vergence anomaly[7] have also been observed. Moreover, many studies have found that an excessive number of fixations of too long duration generally perturb dyslexics in reading, but not in visual tasks[8–10]. Similar results have also been observed in languages with a higher grapheme-phoneme correspondence such as German[11–13] and even in logographic languages with a deep orthography such as Chinese[14]. Indeed, reading requires additional linguistic and cognitive processing demands. However, the saccadic patterns observed in readers with dyslexia seem to be the result and not the cause of their reading disabilities[2,15].

This excess of fixations in dyslexia has recently been studied by many groups[2,5,9,16–20] using different eye-tracking systems available today[21]. This widely observed symptom of dyslexia seems to be able to detect and predict dyslexia using machine learning based on the eye-tracking technique[22–28]. The role of external crowding[29,30] has been studied and discussed in developmental dyslexia[31,32]. Various remediation methods including increasing letter spacing[33] , use of colour filters[34,35], and e-reading with spaced letters[36] have been shown to help people with dyslexia.

A possible role for the lack of asymmetry between the Maxwell centroids in dyslexia inducing an absence of ocular dominance and the frequent existence of perceived extra mirror or duplicated images has also been proposed[37]. The associated internal visual crowding due to callosal interhemispheric projections of letters and words can perturb the brain connectivity[38], in particular in the reading process. It is the aim of this paper to show the role of this internal visual crowding in eye movements, especially in eye fixations during reading. The higher number of fixations being an undisputed symptom of dyslexia, it is tempting to try to control the fixations using the Hebbian mechanisms[39] at the synapses of the primary cortex. To investigate this possibility, we have electronically modified a computer screen equipped with an eye tracker so as to optimize the lighting regime able to control the internal visual crowding. The presence or absence of this internal visual crowding could then worsen or improve the fixational movements and suggests a causal relationship[40] with the reading deficits.

## Methods

### 1- Participants

We tested two students (21 years old) following the same physics courses at the University. The student with dyslexia and the second student with normal reading were aware of the purpose of the study and gave informed written consent before participating. The entire investigation process was conducted according to the principles expressed in the Declaration of Helsinki.

### 2- Foveascope

The setup described in ref 37 is dedicated to investigate the two Maxwell centroid profiles, i.e. the blue cone-free areas at the centre of the foveas and to record their asymmetry (Fig. 1). The contrast of the Maxwell centroid entoptic image is optimized by using a blue-green exchange filter. Each observer adjusts the modulation frequency to his best convenience around 0.2 Hz.

### a) Normal reader

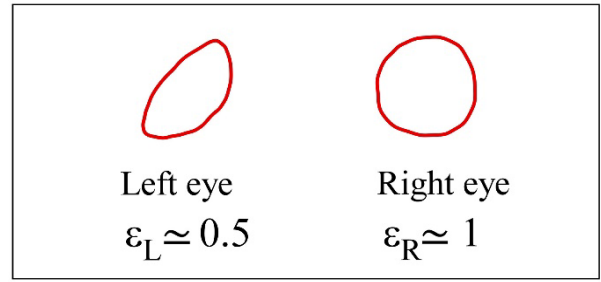

### b) Dyslexic reader

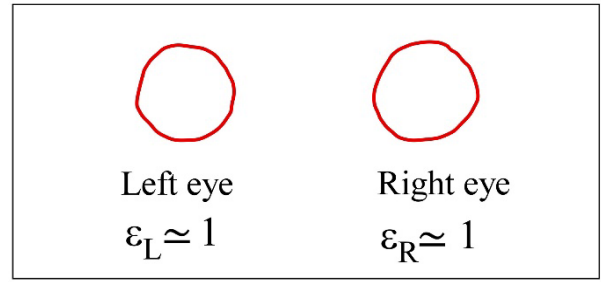

**Figure 1**: Maxwell's centroid profiles.

a) For a normal reader. b) For a dyslexic reader. The profiles show the asymmetry for the normal reader and the lack of asymmetry for the dyslexic reader. The corresponding mean diameters on the retinas are between 100 to 150 μm.

### 3- Noise-activated negative afterimages

Retinal neurons are non-linear and bistable, and therefore sensitive to noise[41]. Here, the closed eyelids allow 2% of the incident light to pass through. This diffuse light constitutes noise falling on the retina that can activate the retinal cells and the primary images arriving on layer 4 of the primary cortex, which is the only layer sensitive to diffuse light[42] and which receives most of the signals from the retinas. After fixating for a few seconds a stimulus (Fig. 2a) such as the word "NEURONS" placed on a window illuminated by daylight, closing the eyes and blocking out all light with the hands placed over the eyes and then shifting them periodically apart, the observer perceives the negative afterimage of the stimulus, as shown in Fig. 2b for the dyslexic reader and in Fig. 2c for the normal reader.

a) Stimulus

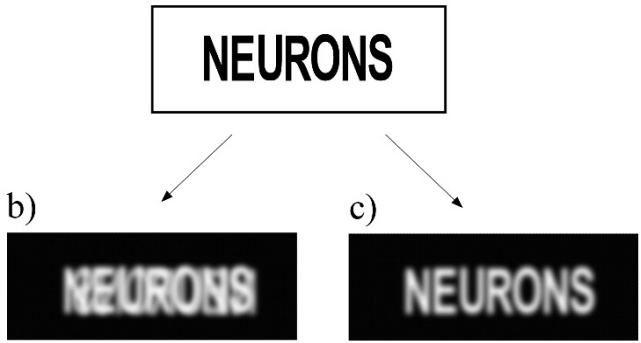

**Figure 2**: Noise-activated negative afterimages.

a) Stimulus: NEURONS. b) Noise-activated negative afterimage perceived by the dyslexic. The mirror-image induces an internal visual crowding. c) Noise-activated negative afterimage perceived by the normal reader.

## 4- Eye tracking movements with an electronically modified computer screen

The eye tracker is a commercial infrared system (tobii dynavox PCEye Plus) with a sampling frequency of 60 Hz. The computer screen was electronically modified so as to be able to work in the continuous lighting regime or in a pulse-width modulated regime with a variable frequency from 60 to 120 Hz. The experiment is carried out in a dark room. Fig. 3a shows the whole system with the corresponding screen luminance recorded in continuous (left side of Fig. 3b) and pulsed regime (right side of Fig. 3b). The mean luminance is the same in both regimes. In the pulsed regime, the cyclic ratio can be adjusted continuously.

## Results

### 1- The asymmetry of Maxwell's centroids

The two students have recorded the profiles of their two Maxwell's centroids shown in Fig. 1. The ellipticity of each profile $\varepsilon_R$ and $\varepsilon_L$ for the right and left eye respectively is measured thanks to the osculating ellipse. The asymmetry is defined by $\Delta\varepsilon = \varepsilon_R - \varepsilon_L$. Here for the normal reader the asymmetry equals $\Delta\varepsilon \simeq 0.5$, with a quasi-circular profile in the right eye, corresponding to his dominant eye (Fig. 1a). In contrast, as noted in ref 37, for the dyslexic reader the two profiles are similar (Fig. 1b) and quasi-circular ($\varepsilon_R \simeq \varepsilon_L \simeq 1$) and the lack of asymmetry induces an absence of ocular dominance and an internal visual crowding (Fig. 2b). Note that when the blue cone topographies are different in the two foveas for a normal reader, the green and red cone topographies are also automatically slightly perturbed. The asymmetry induces two slightly different retinal images and the ocular dominance, but also two slightly different retinoptic maps in particular on layer 4 of the primary cortex where virtually all signals from the retinas arrive[43,44].

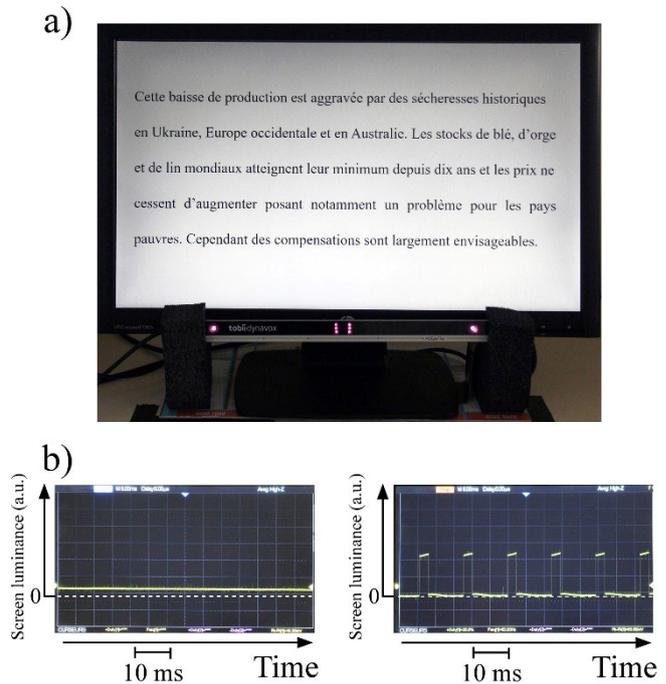

**Figure 3**: a) The eye tracker system with an electronically modified screen. b) Screen luminance versus time in the continuous regime (left side) and in the pulsed regime (right side).

### 2- Internal visual crowding

After a binocular fixation on a stimulus such as NEURONS (Fig. 2a), whereas the normal reader perceived only the primary negative afterimage (Fig. 2c), the dyslexic reader with mirror-images perceived the superposition of the primary and mirror images as in Fig. 2b. Although the mirror–image is weaker, confusion of letters is possible and syllables are difficult to decipher. Mirror-images corresponding to symmetric projections between the two hemispheres were observed in 60 % of a cohort of 160 dyslexic children, whereas duplicated images corresponding to non-symmetric projections were observed in 35 % of the children[45]. As noted previously[37], small lateral shifts of the projected images generally occur leading to different levels of severity of the internal visual crowding.

## a) Dyslexic

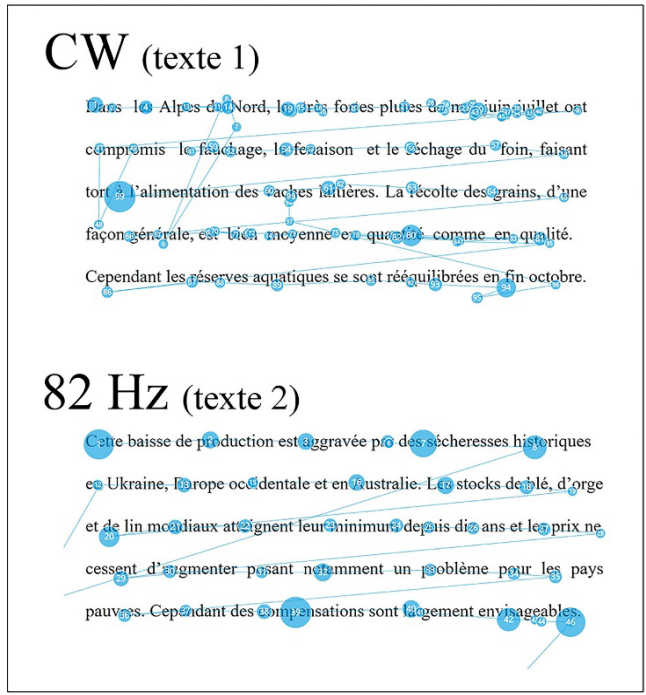

## b) Normal reader

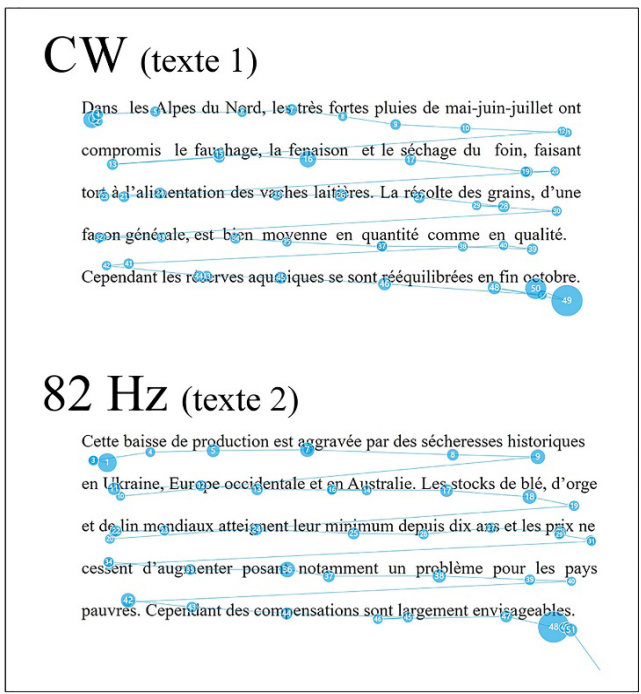

**Figure 4**: Eye movement patterns during reading. a) For a dyslexic in the continuous (top) and pulsed regime (82 Hz – bottom). b) For a control reader in the continuous (top) and pulsed regime (82 Hz – bottom).

### 3- Fixations during reading

The eye movement patterns during reading are shown in Fig. 4 for the two readers under the continuous (CW) and pulsed light regime for two texts. Whereas for the normal reader 50 fixations are necessary whatever the light regime (Fig. 4b), for the dyslexic reader 95 fixations are necessary in the usual continuous regime (top of Fig. 4a), but only 46 in the optimized pulsed light regime at 82 Hz (bottom of Fig. 4a), reaching the normal reader level.

Repeating the experiment for four different similar texts gives the results schematized in Fig. 5. The error bars represent the estimated errors. For the dyslexic (Fig. 5a) the number of fixations is divided by a factor of about 1.8 in the pulsed regime. Without the internal visual crowding, the reader with dyslexia returns to the normal reader level (Fig. 5b).

The total reading times of the two readers are shown in Fig. 6a. While the reading time is invariant for the normal reader, the total time is divided by a factor of about 1.6 in the pulsed regime for the dyslexic reader, but remains longer than that of the normal reader. The fixation durations are shown in Fig. 6b. For both readers the duration times are quasi invariant, but the fixation duration remains longer for the dyslexic reader by about 30 %.

## Discussion

Our eye tracking experiment confirms that the eye movements of the reader with dyslexia are different from those of a normal reader (Fig. 4). In particular, the dyslexic reader makes more fixations (about twice as many as a normal reader), has longer reading times, and makes longer fixations. Such observations have been made in different languages[11,12,14,28]. However, the causal relationship remains discussed[40]. A lack of asymmetry between the Maxwell centroids of the two foveas has been shown to induce internal visual crowding in many readers with dyslexia[37,45] and postural instabilities[46]. Indeed, the retinal

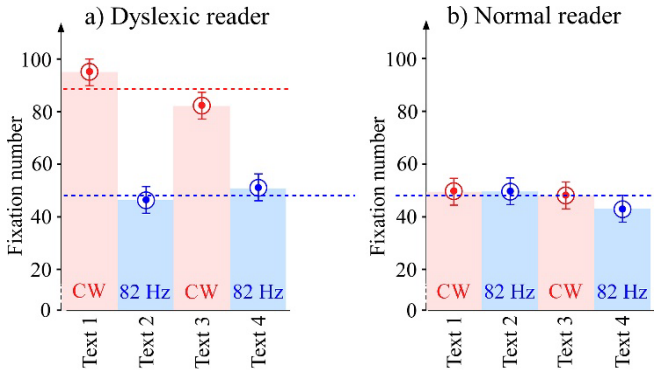

**Figure 5**: Number of fixations for four different texts.

a) For the dyslexic in the two regimes. b) For the normal reader in the two regimes. The red zone corresponds to the continuous regime and the blue zone to the pulsed regime.

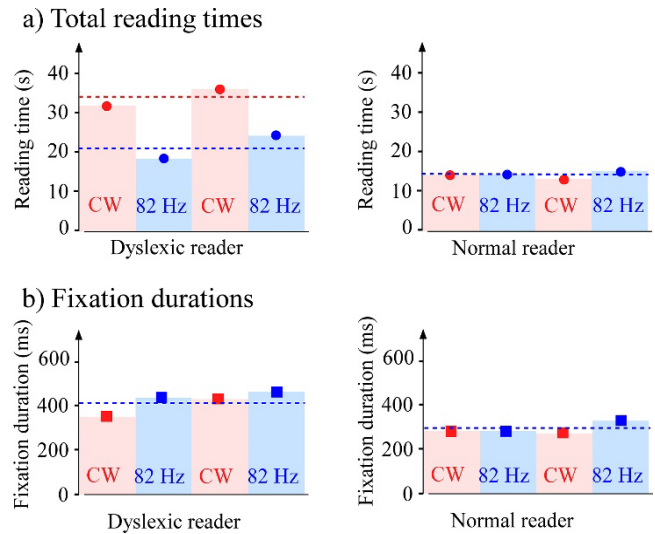

**Figure 6**: Comparative durations for the dyslexic and the control readers.

a)Total duration for four reading texts for the two readers in the two lighting regimes. For the dyslexic reader, at 82 Hz, the total duration is reduced by a factor 1.6 .b) Invariance of the fixation durations for the two readers in the two lighting regimes, but the fixation durations remain longer for the dyslexic reader.

images of the two eyes are too similar, as the two cortical retinoptic neuronal topographies on layer 4 of the primary cortex where the ganglion cells of the retinas reach the cortex. The interhemispheric projections through the corpus callosum between the too similar neuronal topographies in the two hemispheres are stronger than those for a normal reader with an asymmetry. If so, the symmetric projections lead to superposed primary and mirror images and are perceived by the dyslexic reader for letters, but also for words, as shown in Fig. 2b. Internal visual crowding is absent in normal reader (Fig. 2c) and cannot anyway be geometrically weakened by spacing effects like the external crowding which can also induce impairments in reading[31].

In contrast, however, internal visual crowding has been shown to be erasable using the Hebbian mechanisms[39] at the synapses of the primary cortex[37]. Indeed, as the projected mirror-images have to travel through the corpus callosum, they are delayed by about 10 milliseconds corresponding to the transit time between the two hemispheres[47]. Pulse width modulation of the light of the computer screen, at frequencies beyond the visible flicker, allows then the mirror-images to be weakened, restoring a single primary image as perceived by a normal reader. When the modulation frequency is optimized for a given dyslexic reader (here at 82 Hz), the internal visual crowding is really erased and the number of fixations is immediately reduced recovering the normal reader regime. The responses of the normal reader remain invariant whatever the light regime as there is no internal visual crowding (see Figs. 5-6). The causality relationship between the internal visual crowding and the number of fixations is objectively established with an immediate and quantitative effect.

To conclude, the lack of asymmetry between the two Maxwell centroids in readers with dyslexia, which results in a lack of ocular dominance and the existence of an internal visual crowding, leads to a greater number of fixations with longer durations during reading. Indeed, the too strong interhemispheric projections induce generally either perceived extra mirror or duplicated images[45], which make reading difficult, as reading requires linguistic and cognitive processing demands, in contrast to other visual tasks. A greater number of fixations are then necessary to reading. Thanks to Hebbian mechanisms at synapses in the primary cortex activated by an optimized pulsed light regime from an electronically modified computer screen, the embarrassing internal crowding can be weakened and excessive fixations controlled so as to regain the level of normal readers. As the method uses common tracking features, we hope that the results will be confirmed by other groups. Eye tracking of the fixations provides an immediate, precise, and objective quantification of the reduction of the number of fixations in reading and suggests a causality relationship between the reading deficit and internal visual crowding.

# References

1. Martinez-Conde, S., Macknik, S. L. & Hubel, D. H. The role of fixational eye movements in visual perception. *Nat Rev Neurosci* **5**, 229–240 (2004).

2. Rayner, K. Eye movements in reading and information processing: 20 years of research. *Psychological Bulletin* **124**, 372–422 (1998).

3. Rucci, M., Iovin, R., Poletti, M. & Santini, F. Miniature eye movements enhance fine spatial detail. *Nature* **447**, 852–855 (2007).

4. Fenk, L. M. *et al.* Muscles that move the retina augment compound eye vision in Drosophila. *Nature* **612**, 116–122 (2022).

5. Pavlidis, G. Eye Movements in Dyslexia. *Journal of learning disabilities* **18**, 42–50 (1985).

6. Kirkby, J. A., Blythe, H. I., Drieghe, D. & Liversedge, S. P. Reading Text Increases Binocular Disparity in Dyslexic Children. *PLOS ONE* **6**, e27105 (2011).

7. Ghassemi, E. & Kapoula, Z. Is poor coordination of saccades in dyslexics a consequence of reading difficulties? A study case. *Journal of Eye Movement Research* **6**, (2013).

8. Adler-Grinberg, D. & Stark, L. Eye Movements, Scanpaths, and Dyslexia. *Optometry and Vision Science* **55**, 557 (1978).

9. Prado, C., Dubois, M. & Valdois, S. The eye movements of dyslexic children during reading and visual search: impact of the visual attention span. *Vision Res* **47**, 2521–2530 (2007).

10. Bucci, M. P., Nassibi, N., Gerard, C.-L., Bui-Quoc, E. & Seassau, M. Immaturity of the oculomotor saccade and vergence interaction in dyslexic children: evidence from a reading and visual search study. *PLoS One* **7**, e33458 (2012).

11. Hutzler, F., Kronbichler, M., Jacobs, A. M. & Wimmer, H. Perhaps correlational but not causal: No effect of dyslexic readers' magnocellular system on their eye movements during reading. *Neuropsychologia* **44**, 637–648 (2006).

12. Trauzettel-Klosinski, S. *et al.* Eye movements in German-speaking children with and without dyslexia when reading aloud. *Acta Ophthalmol* **88**, 681–691 (2010).

13. Fischer, B. & Hartnegg, K. Instability of Fixation in Dyslexia: Development - Deficits - Training. *Optom Vis Dev* **40**, (2009).

14. Zhao, J., Liu, M., Liu, H. & Huang, C. The visual attention span deficit in Chinese children with reading fluency difficulty. *Res Dev Disabil* **73**, 76–86 (2018).

15. American Academy of Pediatrics, S. on O., Council on Children with Disabilities, American Academy of Ophthalmology, American Association for Pediatric Ophthalmology and Strabismus, & American Association of Certified Orthoptists. Learning Disabilities, Dyslexia, and Vision. *Pediatrics* **124**, 837–844 (2009).

16. De Luca, M., Di Pace, E., Judica, A., Spinell, D. & Zoccolotti, P. Eye movement patterns in linguistic and non-linguistic tasks in developmental surface dyslexia. *Neuropsychologia* **37**, 1407–1420 (1999).

17. Lennerstrand, G., Ygge, J. & Jacobsson, C. Control of Binocular Eye Movements in Normals and Dyslexics. *Annals of the New York Academy of Sciences* **682**, 231–239 (1993).

18. Bucci, M. P. Visual training could be useful for improving reading capabilities in dyslexia. *Appl Neuropsychol Child* **10**, 199–208 (2021).

19. Bonifacci, P., Tobia, V., Sansavini, A. & Guarini, A. Eye-Movements in a Text Reading Task: A Comparison of Preterm Children, Children with Dyslexia and Typical Readers. *Brain Sciences* **13**, 425 (2023).

20. Vagge, A., Cavanna, M., Traverso, C. & Iester, M. Evaluation of ocular movements in patients with dyslexia. *Annals of dyslexia* **65**, (2015).

21. Holmqvist, K. *et al.* Eye tracking: empirical foundations for a minimal reporting guideline. *Behav Res Methods* **55**, 364–416 (2023).

22. Rello, L. & Ballesteros, M. Detecting readers with dyslexia using machine learning with eye tracking measures. in *Proceedings of the 12th International Web for All Conference* 1–8 (Association for Computing Machinery, 2015). doi:10.1145/2745555.2746644.

23. Benfatto, M. N. *et al.* Screening for Dyslexia Using Eye Tracking during Reading. *PLOS ONE* **11**, e0165508 (2016).

24. Asvestopoulou, T. *et al.* DysLexML: Screening Tool for Dyslexia Using Machine Learning. Preprint at https://doi.org/10.48550/arXiv.1903.06274 (2019).

25. Raatikainen, P. *et al.* Detection of developmental dyslexia with machine learning using eye movement data. *Array* **12**, 100087 (2021).

26. El Hmimdi, A. E., Ward, L. M., Palpanas, T. & Kapoula, Z. Predicting Dyslexia and Reading Speed in Adolescents from Eye Movements in Reading and Non-Reading Tasks: A Machine Learning Approach. *Brain Sci* **11**, 1337 (2021).

27. JothiPrabha, A., Bhargavi, R. & Deepa Rani, B. V. Prediction of dyslexia severity levels from fixation and saccadic eye movement using machine learning. *Biomedical Signal Processing and Control* **79**, 104094 (2023).

28. Vajs, I., Papić, T.,Ković, V., Savić, A. M. & Janković, M. M. Accessible Dyslexia Detection with Real-Time Reading Feedback through Robust Interpretable Eye-Tracking Features. *Brain Sciences* **13**, 405 (2023).

29. Pelli, D. G. Crowding: a cortical constraint on object recognition. *Current Opinion in Neurobiology* **18**, 445–451 (2008).

30. Levi, D. M. Crowding—An essential bottleneck for object recognition: A mini-review. *Vision Research* **48**, 635–654 (2008).

31. Martelli, M., Di Filippo, G., Spinelli, D. & Zoccolotti, P. Crowding, reading, and developmental dyslexia. *Journal of Vision* **9**, 14 (2009).

32. Bellocchi, S. Developmental Dyslexia, Visual Crowding and Eye Movements. in *Eye movements: Developmental perspectives, dysfunctions and disorders in humans.* (Nova Science Publishers, Ed. Steward L.C., 2013).

33. Bertoni, S., Franceschini, S., Ronconi, L., Gori, S. & Facoetti, A. Is excessive visual crowding causally linked to developmental dyslexia? *Neuropsychologia* **130**, 107–117 (2019).

34. Hall, R., Ray, N., Harries, P. & Stein, J. A comparison of two-coloured filter systems for treating visual reading difficulties. *Disabil Rehabil* **35**, 2221–2226 (2013).

35. Razuk, M. *et al.* Effect of colored filters on reading capabilities in dyslexic children. *Research in Developmental Disabilities* **83**, 1–7 (2018).

36. Schneps, M. H., Thomson, J. M., Chen, C., Sonnert, G. & Pomplun, M. E-Readers Are More Effective than Paper for Some with Dyslexia. *PLOS ONE* **8**, e75634 (2013).

37. Le Floch, A. & Ropars, G. Left-right asymmetry of the Maxwell spot centroids in adults without and with dyslexia. *Proc R Soc B* **284**, 20171380 (2017).

38. Finn, E. S. *et al.* Disruption of Functional Networks in Dyslexia: A Whole-Brain, Data-Driven Analysis of Connectivity. *Biological Psychiatry* **76**, 397–404 (2014).

39. Hebb, D. O. *The organization of behavior; a neuropsychological theory.* (Wiley, 1949).

40. Werth, R. Dyslexia: Causes and Concomitant Impairments. *Brain Sciences* **13**, 472 (2023).

41. Longtin, A., Bulsara, A., Pierson, D. & Moss, F. Bistability and the dynamics of periodically forced sensory neurons. *Biol. Cybern.* **70**, 569–578 (1994).

42. Hubel, D. H. *Eye, brain, and vision.* (Scientific American Library, 1988).

43. Hubel, D. H. & Wiesel, T. N. Ferrier lecture - Functional architecture of macaque monkey visual cortex. *Proceedings of the Royal Society of London. Series B. Biological Sciences* **198**, 1–59 (1997).

44. Crick, F. *Astonishing Hypothesis: The Scientific Search for the Soul.* (Scribner, 1995).

45. Le Floch, A. & Ropars, G. Le manque d'asymétrie des centroïdes de Maxwell, et de dominance oculaire, chez les dyslexiques. *Revue Francophone d'Orthoptie* **13**, 134–138 (2020).

46. Floch, A. L., Henriat, S., Fourage, R. & Ropars, G. Postural Instability in a Young

Dyslexic Adult Improved by Hebbian Pulse-width Modulated Lighting. *American Journal of Internal Medicine* **8**, 267 (2020).

47. Caminiti, R. *et al.* Diameter, Length, Speed, and Conduction Delay of Callosal Axons in Macaque Monkeys and Humans: Comparing Data from Histology and Magnetic Resonance Imaging Diffusion Tractography. *J. Neurosci.* **33**, 14501–14511 (2013).

## Acknowledgments

We thank the two students for their kind participation, the University of Rennes for access to its facilities, and J. R. Thébault for his technical assistance.

## Authors contributions

ALF designed the study and wrote the manuscript. GR and ALF developed the methods and the apparatus and contributed to the interpretation of the data and discussed the results.

## Competing interest

A patent has been filed by the University of Rennes for the modified computer screen.

## Data accessibility

This article has no additional data.